# The Sagnac effect in Coupled-Resonator Slow-Light Waveguide Structures

Jacob Scheuer

Center for the Physics of Information, California Inst. of Technology, M/C 128-95, Pasadena, CA 91125.

koby@caltech.edu

Abstract

We study the effect of rotation on the propagation of electromagnetic waves in slow-light waveguide structures consisting of coupled micro-ring resonators. We show that such configurations exhibit new a type of the Sagnac effect which can be used for the realization of highly-compact integrated rotation sensors and gyroscopes.

When an electromagnetic wave propagates in a moving medium it accumulates additional phase shift, compared to a wave propagating in a stationary medium, which depends on the scalar product between the wave propagation direction and the velocity vector of the medium [1, 2]. A particularly interesting configuration is that of a wave propagating along a circular path in a rotating medium. In such scenario, the additional phase accumulated by the wave depends on the relation between the propagation directions of the medium and the wave (co-directional or counter-directional). This phase difference is often referred to as the Sagnac effect and in addition to its scientific importance, it has numerous practical application such as detection and high-precision measurement of rotation.

In the past few years, much attention was devoted to slowing down the propagation speed of light and to coherently stop and store pulses of light [3-6]. There are two major approaches to achieve significant reduction of the group velocity of light, which employ either electronic or optic resonances. Because of the inherent constraints associated with the conversion of the optical signals to coherent electronic states, the electronic resonance approach is less attractive for practical implementations of slow-light devices. Consequently, significant efforts were focused on controlling the speed of light using photonic structure incorporating microcavities and photonic crystals. Substantial delays and storage of light pulses were predicted in various coupled-cavities structures such as coupled resonator optical waveguides (CROWs) [7] and side-coupled integrated spaced sequence of resonators (SCISSORs) [8].

Recently, Leonhardt et al. pointed out the advantages of using the Sagnac effect is slow-light medium generated by electromagnetically induced transparency (EIT) for the realization of an ultra-sensitive optical gyroscope [9]. Subsequently, Steinberg studied the effect of rotation in coupled photonic crystal defect cavities [10] and Matsko et al. proposed to utilize the dispersive characteristics of slow-light propagation in a closed loop SCISSOR-like configuration to realize a high-sensitivity miniaturized optical gyroscope [11]. In that study, however, the SCISSOR was modeled as a highly-dispersive conventional waveguide where the slow group velocity of the light in the SCISSOR stems from the average interaction of the light with the high-$Q$ resonators.

In this letter, we study the properties of the Sagnac effect in a CROW which is wrapped around itself, with application for a highly compact rotation sensor or an optical gyroscope. Figure 1 illustrates the geometrical configuration: light is launched into the input waveguide and equally divided between the two channels of the 3dB coupler. The signal in each arm is coupled to a different end of the circular CROW consisting of directly coupled ring resonators. Finally, the counter propagating signals (marked by the black and white arrows) are combined by the 3dB coupler where the output signal in each arm of the coupler depends on the relative phase difference between the signals:

$$|A_r|^2 = |E_{in}|^2 \cos^2\left(\tfrac{1}{2}\Delta\phi\right); \; |B_r|^2 = |E_{in}|^2 \sin^2\left(\tfrac{1}{2}\Delta\phi\right) \tag{1}$$

where $E_{in}$ and $\Delta\phi$ are correspondingly the input amplitude and the phase difference between the counter-rotating fields. When the device is stationary, the overall phases accumulated by both signals are identical i.e., $\Delta\phi = 0$, resulting in complete cancellation of $B_r$. On the other hand, when the device is rotating, the phases accumulated by the signals differ, resulting in a non-vanishing intensity $B_r$.

To evaluate the phase difference $\Delta\phi$ in a CROW it is convenient to divide the structure into sections as illustrated in Fig. 1: An input section which consists of the input coupler and part of the first micro-ring (this section is marked by the dashed white line "I"); A recurring section consisting of two halves of a micro-ring coupled to a complete ring, constituting the main body of the CROW (this section is defined by two successive dashed white lines: I→$S_1$, $S_1$→$S_2$, etc.); And an output section which is similar to the input section (from the line marked by "O" to the output coupler). Because of the recurring section, it is convenient to represent each section by a transfer matrix linking between the input and output ports of the section. The overall transfer matrix of the structure is then found simply by multiplying these matrices in the correct order.

The phase accumulated by a wave propagating in non-stationary waveguide depends primarily on the scalar product of the waveguide velocity and the wave-vector $k$. In the configuration studied here, the contribution of each segment $d\vec{r}$ in each micro-ring is different because the center of rotation does not necessarily coincide with the center of any of the micro-rings. Therefore, in order to construct the transfer matrix of each section we have to evaluate the phase accumulated by a wave propagating along a curved waveguide segment which is rotating around an arbitrary point.

Figure 2 illustrate the geometry of this problem: a wave propagating in micro-ring resonator with radius $R$ while the center of this ring is rotating with angular velocity Ω around a fixed point. The distance between the center of the micro-ring and the center of rotation is $\widetilde{R}_0$. The phase accumulated by the wave as it propagates along a segment $d\vec{r}$ stem from two contributions: The conventional phase due to the propagation $d(\Delta\phi_{\text{prop}})=\omega/cn|dr|$ and the *rotation related* phase shift which is given by [1]:

$$d(\Delta\phi_{\text{rot}}) = \frac{\omega n^2}{c^2}(1-\alpha)\vec{V}\cdot d\vec{r} \tag{2}$$

where $\omega$ is the optical (angular) frequency, $c$ is the speed of light in vacuum, $n$ is the refractive index, $\vec{V}=\vec{\Omega}\times(\vec{\widetilde{R}}_0+\vec{R})$ is the linear velocity of the segment, and $\alpha$ is the Fresnel-Fizeau drag coefficient given by $\alpha = c(1\text{-}n^{-2})$ (for a non-dispersive medium). Therefore, the overall rotation-related phase accumulated by an electromagnetic wave which propagates in a micro-ring from $\theta_{\text{s}}$ to $\theta_{\text{f}}$ (see Fig. 2) is given by:

$$\Delta\phi_{\text{rot}} = \int_{\theta_{\text{s}}}^{\theta_{\text{f}}} d(\Delta\phi_{\text{rot}}) = \frac{\omega\Omega}{c^2}R^2(\theta_{\text{f}}-\theta_{\text{s}}) + \frac{\omega\Omega}{c^2}R\widetilde{R}_0(\cos\theta_{\text{f}}-\cos\theta_{\text{s}}) \tag{3}$$

Eq. (3) exhibits several interesting properties that should be noted. First, the rotation-related phase shift is independent of the waveguide index of refraction $n$. This is a well-known property of the Sagnac effect which does not depend on the refractive index of the medium comprising the loop. Second, for a complete loop the second term in (3) vanishes and thus the phase shift is independent of the center of the rotation. However, in the structure analyzed here, the propagation section between two couplers does not form a complete loop, and therefore, the second term must be included.

For simplicity, we assume that the micro-rings are identical and lossless and that the coupling coefficients $\kappa$ between adjacent micro-rings are also identical. The transfer matrices for the three types of section are straightforwardly given by:

$$
\begin{aligned}
M_I &= \begin{pmatrix} 0 & \exp i\left(\phi_P^{\alpha/2} + D\phi_S^{\alpha/2} + \phi_R^{S,\alpha/2}\right) \\ \exp i\left(\phi_P^{\alpha/2-\pi} + D\phi_S^{\alpha/2-\pi} + \phi_R^{S,\alpha/2}\right) & 0 \end{pmatrix} C(\kappa) \\
M_O &= C(\kappa) \begin{pmatrix} \exp i\left(\phi_P^{\alpha/2} + D\phi_S^{\alpha/2} - \phi_R^{S,\alpha/2}\right) & 0 \\ 0 & \exp i\left(\phi_P^{\alpha/2-\pi} + D\phi_S^{\alpha/2-\pi} - \phi_R^{S,\alpha/2}\right) \end{pmatrix} \\
M_R &= -M_C C(\kappa) \begin{pmatrix} \exp i\left(\phi_P^{-(\pi+\alpha)} + D\phi_S^{(\pi+\alpha)} + \phi_R^{C,\alpha}\right) & 0 \\ 0 & \exp i\left(\phi_P^{-(\alpha-\pi)} + D\phi_S^{(\alpha-\pi)} + \phi_R^{C,\alpha}\right) \end{pmatrix} C(\kappa) M_C \\
M_C &= \begin{pmatrix} \exp i\left(\phi_P^{(\pi+\alpha)/2} + D\phi_S^{(\pi+\alpha)/2} + \phi_R^{C,\alpha/2}\right) & 0 \\ 0 & \exp i\left(\phi_P^{(\alpha-\pi)/2} + D\phi_S^{(\alpha-\pi)/2} + \phi_R^{C,\alpha/2}\right) \end{pmatrix} \\
C(\kappa) &= \frac{i}{\sqrt{\kappa}} \begin{pmatrix} \sqrt{1-\kappa} & -1 \\ 1 & -\sqrt{1-\kappa} \end{pmatrix}
\end{aligned} \tag{4}
$$

$$\phi_P^{\theta} = \omega n R\theta/c;\ \ \phi_S^{\theta} = \omega\Omega R^2\theta/c^2;\ \phi_R^{S,\theta} = \omega\Omega R\tilde{R}_0 \sin(\theta)/c^2;\ \phi_R^{C,\theta} = \omega\Omega R\tilde{R}_0 \cos(\theta)/c^2$$

where $M_I$, $M_O$, and $M_R$ are respectively the transfer matrices for the input section, the output section and the recurring section, $D$=1 for the signal propagating with the device rotation and –1 for the signal counter propagating to the device rotation. $\alpha$ is the angle between adjacent micro-rings (see Fig. 1). The overall transfer matrix connecting between the inputs and outputs of the CROW is, therefore, given by:

$$M_{\mathrm{CROW}} = M_O \left(M_R\right)^{(N-1)/2} M_I \tag{5}$$

where $N$ is the number of micro-ring comprising the CROW which must be odd for the configuration illustrated in Fig. 1.

Equations (5), (4) and (1) allow us to calculate the output signal $B_r$ for various parameters. A closer inspection of (4) allows us to eliminate some of the terms because we are not interested in the complete transmission function of the CROW but rather in the phase difference between the two paths. The phase terms proportional to $\tilde{R}_0$ can be rewritten as unit matrices multiplied by a common phase factor. Since this phase factor is identical for both paths, it has no effect on the outcome of (1), and therefore, the output signal $B_r$ is independent of $\tilde{R}_0$. This is an important conclusion because $\tilde{R}_0$ defines the area of the effective CROW ring. The Sagnac effect in conventional waveguide loops is directly related to the area of the loop, thus we cannot analyze this effect in the CROW loop simply by assuming an effective ring waveguide with the dispersion relation of a CROW.

Fig. 3a depicts the output intensity $|B_r|^2$ as a function of $\Omega$ for CROWs with different number of resonators. The parameters of these CROWs are defined in the figure caption. As can be expected, the output intensity increases for larger $\Omega$ with steeper slope (responsivity) for larger number of micro-rings. For rotation-sensing application, a steeper slope is advantageous because it corresponds to higher sensitivity, i.e., ability to detect slower rotation rates. Figures 3b-3d show the relative responsivity of the CROW loop for varying number of rings (3b), coupling coefficient (3c) and the micro-rings radius (3d). The responsivity increases for larger number of micro-rings $N$, smaller coupling coefficient $\kappa$, and larger micro-rings radius $R$.

It is worthy to quantify some of these trends because they reveal the inherent differences between the Sagnac effect in CROWs and in conventional waveguides. Figure 3b shows a quadratic fit to the dependence of the responsivity on the number of rings comprising the CROW. The fit indicates that the responsivity of a closed-loop CROW, $S_N$, consisting of $N$ micro-rings is related to that of a single ring according to $S_N = (N+1)^2/4 \cdot S_1$. It is well known that the slope of the output signal of a single-ring device is proportional to the square of the ring area [12], and therefore, the responsivity of the CROW-based device is proportional to the square of the total area of the micro-rings composing it. It should be emphasized that, unlike the conventional Sagnac effect, the overall area circumscribed by the coupled-resonator waveguide does not affect the output signal. This result, which clearly demonstrates the difference between the Sagnac effect in conventional and in CROWs, is interesting and, to some extent, counter-intuitive because one might expect the Sagnac effect contributions from adjacent micro-rings to cancel each other. Figure 3c also compares between the numerically calculated responsivity according to (5) and an analytic expression derived according to the responsivity of a single micro-ring and the quadratic dependence of the responsivity on the number of micro-rings in the CROW.

For practical applications, the CROW-based gyroscope exhibits several inherent advantages compared to conventional Sagnac loops: 1) The dependence of the gyro output signal on the inter-ring coupling allows to improve the device sensitivity without requiring larger area; 2) The independence of the responsivity of the CROW-gyro on $\tilde{R}_0$ indicates that the arrangement of micro-rings comprising the CROW is insignificant, and thus, enabling a more efficient utilization of the chip area.

The limiting factor of the ability of a rotation sensor to detect low angular velocity is the output power $|B_r|^2$ compared to the Shot noise. While ideal micro-ring resonators are lossless, when light propagates in real resonators it experiences propagation loss that can be introduced into our analysis by introducing an imaginary part to the index of refraction in (4). The propagation loss decreases the output signal and reduces the attainable sensitivity of the rotation sensor. Figure 4 shows the responsivity of a CROW rotation sensor as a function of the resonators' Quality-factor ($Q$). As shown in the figure, for resonators with $Q>10^7$, the influence of the propagation loss is negligible and has small effect on the device response. Since high-$Q$ ($>10^7$) single-mode, planar-technology-based micro-ring resonators are being fabricated by many research groups [13], the propagation losses in the cavities do not limit significantly the sensitivity of the CROW rotation sensor.

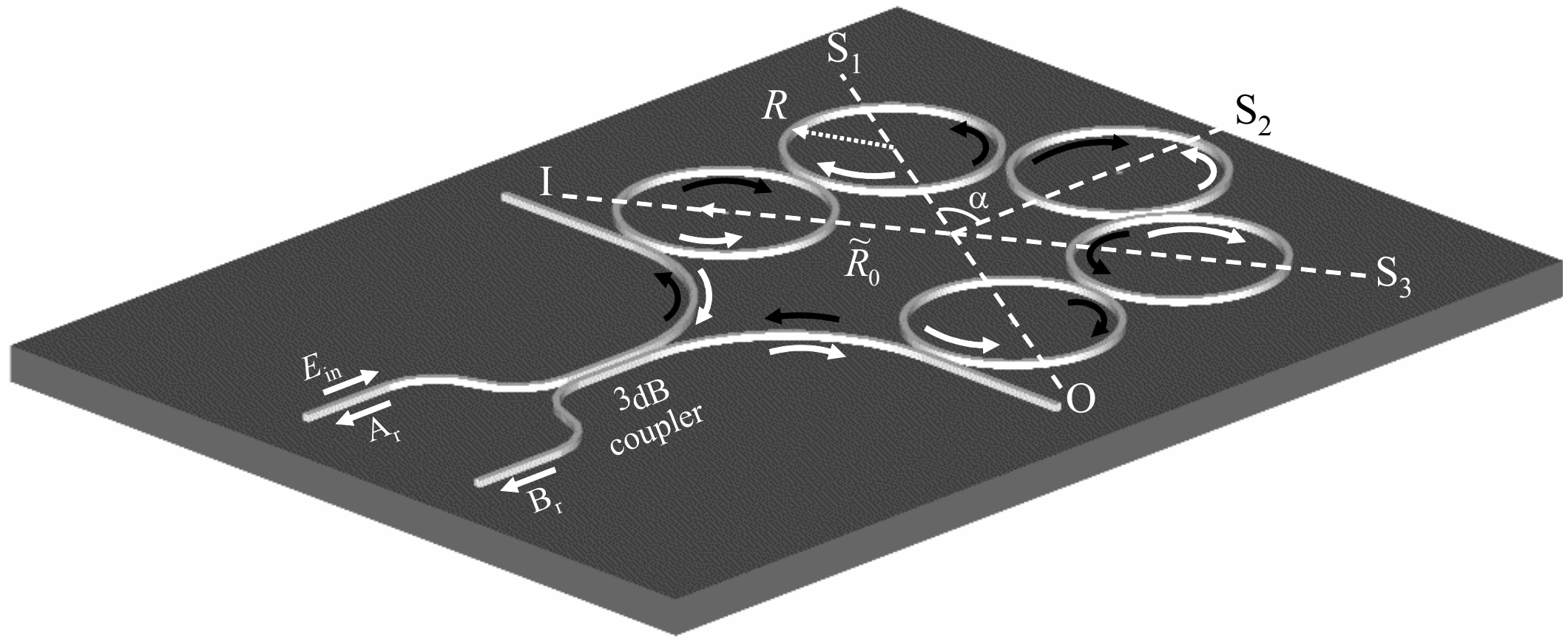

Fig. 1. Schematic of the Coupled Resonator Slow-Light rotation sensor

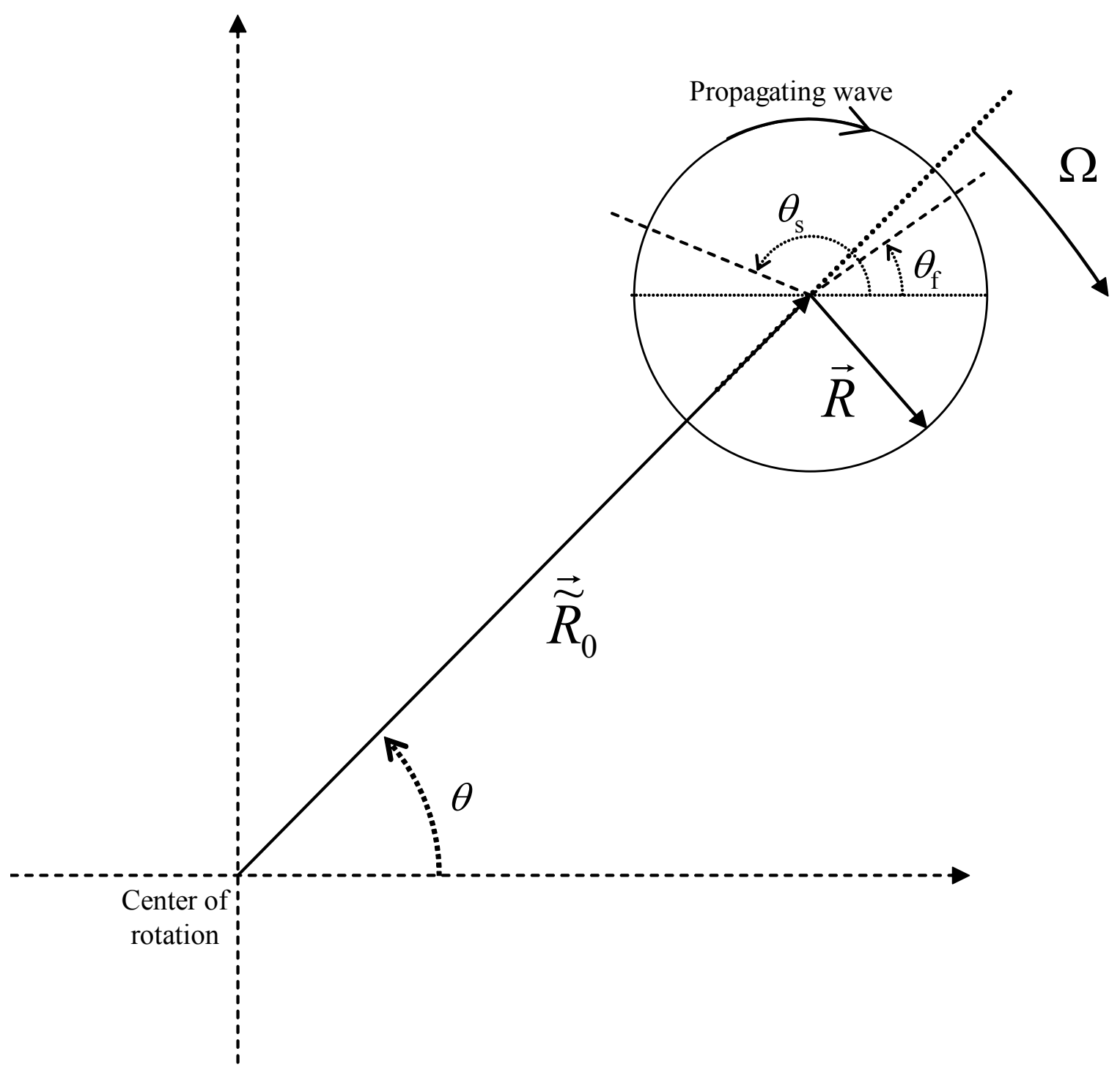

Fig. 2. Phase accumulation in a rotating micro-ring resonator

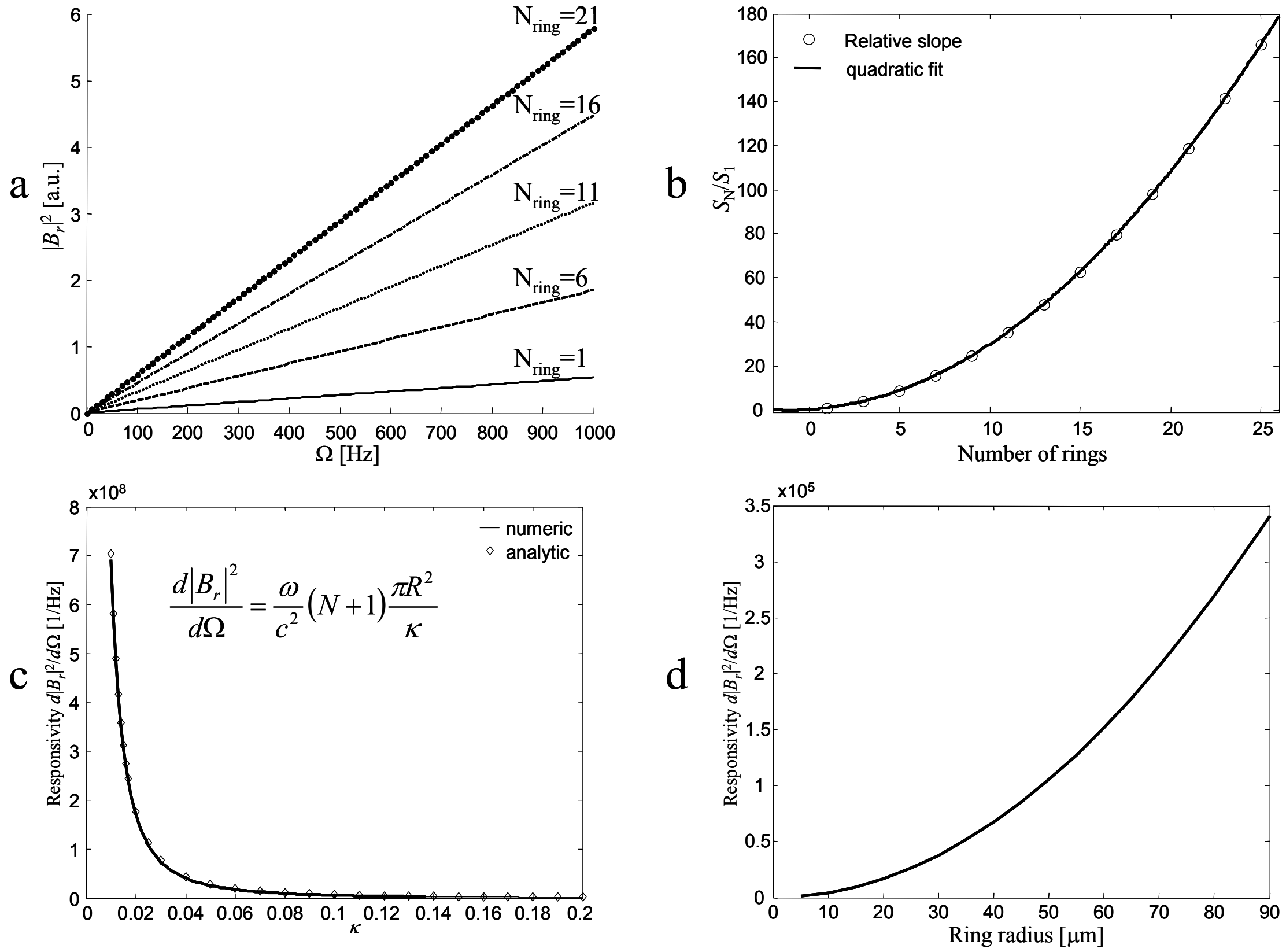

Fig. 3. (a) Output signal intensity as a function of the structure angular velocity for various number of rings, $R$=25μm, $\kappa$=0.01; (b) Dependence of the relative sensitivity on the number of micro-rings, $R$=25μm, κ=0.01; Dependence of the sensitivity on: (c) the coupling coefficient ($R$=25μm, $N$=9) and on (d) the micro-rings radius ($N$=9, $\kappa$=0.01).

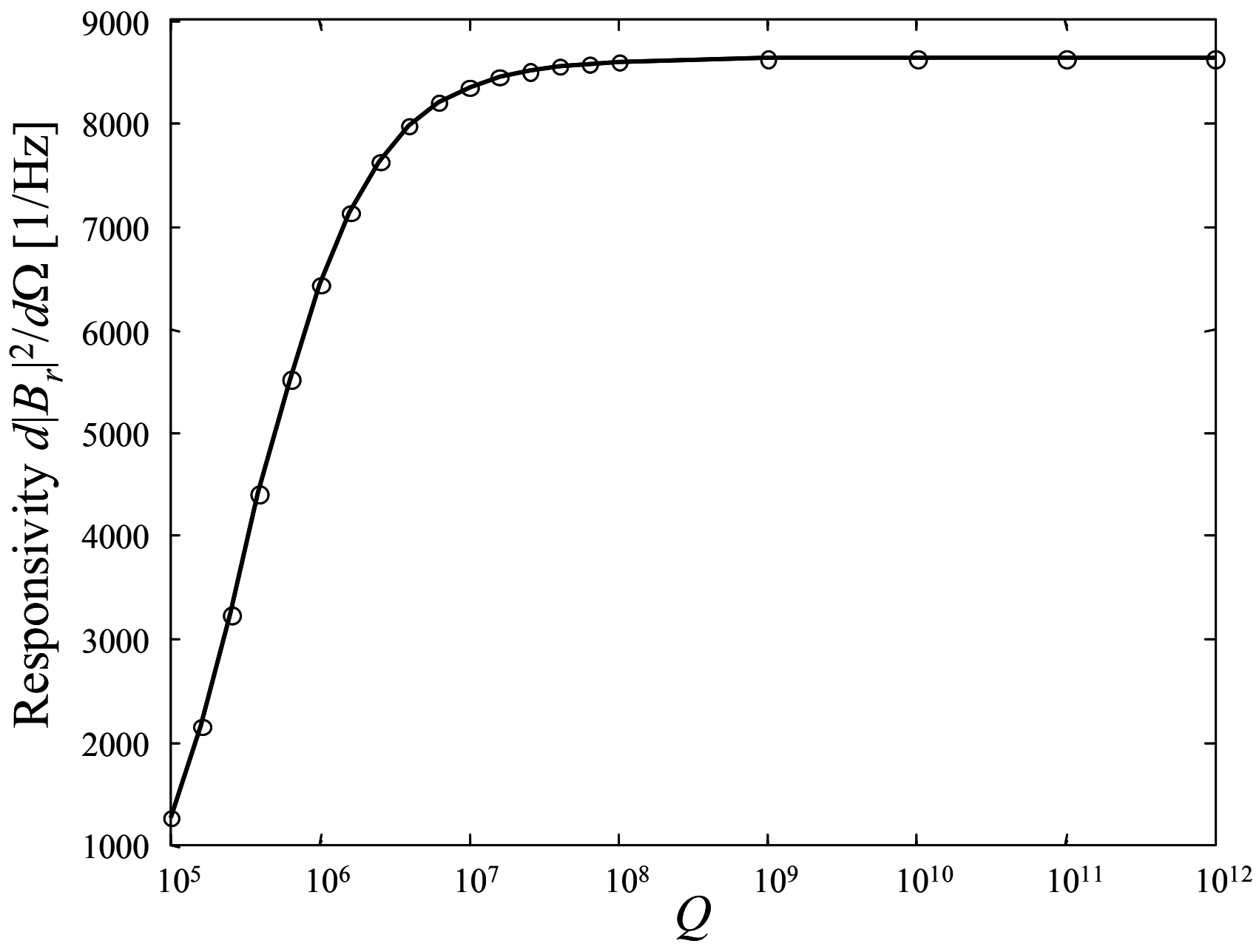

Fig. 4. The impact of the Quality-factor on the sensitivity. $R$=25μm, $N$=9, $\kappa$=0.03.